\documentclass[a4paper]{article}

\usepackage{makecell}
\usepackage{xcolor}
\usepackage{url}
\usepackage{multirow}
\usepackage{INTERSPEECH2019}
\def\newpara{\vspace{7pt}}
\newcommand{\ms}{\mathbf{S}}

\newcommand\blfootnote[1]{%
  \begingroup
  \renewcommand\thefootnote{}\footnote{#1}%
  \addtocounter{footnote}{-1}%
  \endgroup
}

\title{In defence of metric learning for speaker recognition}
\name{Joon Son Chung, Jaesung Huh, Seongkyu Mun, Minjae Lee, Hee Soo Heo, \\
Soyeon Choe, Chiheon Ham, Sunghwan Jung, Bong-Jin Lee, Icksang Han }
\address{Naver Corporation, South Korea}
\email{joonson.chung@navercorp.com}

\begin{document}

\maketitle
\begin{abstract}
The objective of this paper is `open-set' speaker recognition of unseen speakers, where ideal embeddings should be able to condense information into a compact utterance-level
representation that has small intra-speaker and large inter-speaker distance. 

A popular belief in speaker recognition is that networks trained with classification objectives outperform metric learning methods. 
In this paper, we present an extensive evaluation of most popular loss functions for speaker recognition on the VoxCeleb dataset. We demonstrate that the vanilla triplet loss shows competitive performance compared to classification-based losses, and those trained with our proposed metric learning objective outperform state-of-the-art methods. \\

\end{abstract}

\noindent\textbf{Index Terms}: speaker recognition, speaker verification, metric learning.


\section{Introduction}
\blfootnote{\hspace{-12pt}The code for this paper can be found at:\\ \url{https://github.com/clovaai/voxceleb_trainer}}
Research on speaker recognition has a long history and has received
an increasing amount of attention in recent years.
Large-scale datasets for speaker recognition such as the VoxCeleb~\cite{Nagrani17,Chung18a} and Speakers in the Wild~\cite{McLaren16} have become freely available, facilitating fast progress in the field.

Speaker recognition can be categorised into closed-set or open-set settings. For closed-set setting, all testing identities are predefined in training set, therefore can be addressed as a classification problem. For open-set setting, the testing identities are not seen during training, which is close to practice. This is a metric learning problem in which voices must be mapped to a discriminative embedding space. The focus of this research, and most others, are on the latter problem.

Pioneering work on speaker recognition using deep neural networks have learnt speaker embeddings via the classification loss~\cite{Nagrani17,snyder2017deep,snyder2018x}.
Since then, the prevailing method has been to use softmax classifiers to train the embeddings~\cite{ravanelli2018speaker,okabe2018attentive,snyder2019speaker}. 
While the softmax loss can learn separable embeddings, they are not discriminative enough since it is not explicitly designed to optimise embedding similarity.  
Therefore, softmax-trained models have often been combined with PLDA~\cite{Ioffe06} back-ends to generate scoring functions~\cite{snyder2018x,ramoji2020pairwise}. 

This weakness has been addressed by \cite{liu2017sphereface} who have proposed angular softmax (A-Softmax) where cosine similarity is used as logit input to the softmax layer, and a number of works have demonstrated its superiority over vanilla softmax in speaker recognition~\cite{ravanelli2018speaker,okabe2018attentive,snyder2019speaker,villalba2019state,snyder2019jhu}. 
Additive margin variants, AM-Softmax~\cite{wang2018additive,wang2018cosface} and AAM-Softmax~\cite{deng2019arcface}, have been proposed to increase inter-class variance by introducing a cosine margin penalty to the target logit, and these have been very popular due to their ease of implementation and good performance~\cite{Xie19a,hajibabaei2018unified,liu2019large,garcia2019x,zeinali2019but,luu2019channel,luu2020dropclass,xiang2019margin}. However, training with AM-Softmax and AAM-Softmax has proven to be challenging since they are sensitive to the value of scale and margin in the loss function.

Metric learning objectives present strong alternatives to the prevailing classification-based methods, by learning embeddings directly.
Since open-set speaker recognition is essentially a metric learning problem, 
the key is to learn features that have small intra-class and large inter-class distance. 
Contrastive loss~\cite{Chopra05} and triplet loss~\cite{Schroff15} have been demonstrated promising performance on speaker recognition~\cite{zhang2018text,rahman2018attention} by optimising the distance metrics directly, but these methods require careful pair or triplet selection which can be time consuming and performance sensitive.

Of closest relevance to our work is
prototypical networks~\cite{snell2017prototypical} that learn a metric space in which open-set classification can be performed by computing distances to prototype representations of each class, with a training procedure that mimics the test scenario. The use of multiple negatives helps to stabilise learning since loss functions can enforce that an embedding is far from all negatives in a batch, rather than one particular negative in the case of triplet loss. \cite{wang2019centroid,anand2019few} have adopted the prototypical framework for speaker recognition.
Generalised end-to-end loss~\cite{wan2018generalized}, originally proposed for speaker recognition, is also closely related to this setup.

Comparing different loss functions from prior works can be challenging and unreliable, since speaker recognition systems can vary widely in their design. Popular trunk architectures include TDNN-based systems such as x-vector~\cite{snyder2018x} and its deeper counterparts~\cite{snyder2019speaker}, as well as network architectures from the computer vision community such as the ResNet~\cite{He16}. A range of encoders have been proposed to aggregate frame-level informations into utterance-level embeddings, from simple averaging~\cite{Nagrani17} to statistical pooling~\cite{snyder2017deep,okabe2018attentive} and dictionary-based encodings~\cite{Xie19a,cai2018exploring}. 
\cite{snyder2018x} has proven that data augmentation can significantly boost speaker recognition performance, but the augmentation methods can range from adding noise~\cite{snyder2015musan} to room impulse response (RIR) simulation~\cite{allen1979image}.

Therefore, in order to directly compare a range of loss functions, we conduct over 20,000 GPU-hours of careful experiments while keeping other training details constant. Against popular belief, we demonstrate that the networks trained with vanilla triplet loss show competitive performance compared to most AM-Softmax and AAM-Softmax trained networks, and those trained with our proposed angular objective outperform all comparable methods.


\section{Training functions}
\label{sec:loss}

This section describes the loss functions used in our experiments, including a new angular variant of the prototypical loss.

\subsection{Classification objectives}

The VoxCeleb2 development set contains $C=5,994$ speakers or classes. During training, each mini-batch contains $N$ utterances each from different speakers, whose embeddings are $\mathbf{x}_{i}$ and the corresponding speaker labels are $y_{i}$ where $1 \leq i \leq N$ and $1 \leq y \leq C$.

\newpara\noindent\textbf{Softmax.}
The softmax loss consists of a softmax function followed by a multi-class cross-entropy loss. It is formulated as:
\begin{equation}
\footnotesize
L_\text{S}=-\frac{1}{N}\sum_{i=1}^N\log \frac{e^{\mathbf{W}^T_{y_i}\mathbf{x}_i+b_{y_i}}}{\sum_{j=1}^C e^{\mathbf{W}^T_{j}\mathbf{x}_i+b_{j}}}
\end{equation}  
where $\mathbf{W}$ and $b$ are the weights and bias of the last layer of the trunk architecture, respectively.  
This loss function only penalises classification error, and does not explicitly enforce intra-class compactness and inter-class separation.

\newpara\noindent\textbf{AM-Softmax (CosFace).}
By normalising the weights and the input vectors, softmax loss can be reformulated such that
the posterior probability only relies on cosine of angle between the weights and the input vectors. This loss function, termed by the authors as Normalised Softmax Loss (NSL), is formulated as:
\begin{equation}
\footnotesize
L_\text{N}=-\frac{1}{N}\sum_{i=1}^N\log \frac{e^{{\cos(\theta_{{y_i}, i}})}}{\sum_{j} e^{{\cos(\theta_{{j}, i}})}}
\end{equation}
where $\cos{(\theta_{j,i})}$ is the dot product of normalised vector $\mathbf{W}_j$ and $\mathbf{x}_i$.

However, embeddings learned by the NSL are not sufficiently discriminative because the NSL only penalises classification error. In order to mitigate this problem, cosine margin $m$ is incorporated into the equation:
\begin{equation}
\footnotesize
L_\text{C}=-\frac{1}{N}\sum_{i=1}^N\log \frac{e^{s({\cos(\theta_{{y_i}, i})-m})}}{e^{s({\cos(\theta_{{y_i}, i})-m)}} + \sum_{j\neq{y_i}} e^{s({\cos(\theta_{{j}, i})})}}
\end{equation}
where $s$ is a fixed scale factor to prevent gradient from getting too small in training phase.

\newpara\noindent\textbf{AAM-Softmax (ArcFace).} This is equivalent to CosFace except that there is additive {\em angular} margin penalty $m$ between $\mathbf{x}_i$ and $\mathbf{W}_{y_i}$. The additive angular margin penalty is equal to the geodesic distance margin penalty in the normalised hypersphere.
\begin{equation}
\footnotesize
L_\text{A}=-\frac{1}{N}\sum_{i=1}^N\log \frac{e^{s({\cos(\theta_{{y_i}, i}+m)})}}{{e^{s({\cos(\theta_{{y_i}, i}+m))}} + \sum_{j\neq{y_i}} e^{s({\cos(\theta_{{j}, i})})}}}
\end{equation}

\subsection{Metric learning objectives}

For metric learning objectives, each mini-batch contains $M$ utterances from each of $N$ different speakers, whose embeddings are $\mathbf{x}_{j,i}$ where $1 \leq j \leq N$ and $1 \leq i \leq M$.

\newpara\noindent\textbf{Triplet.} Triplet loss minimises the $L2$ distance between an anchor and a positive (same identity), and maximises the distance between an anchor and a negative (different identity). 
\begin{equation}
\footnotesize
L_\text{T}= \frac{1}{N}\sum_{j=1}^N\text{max}(0,\lVert \mathbf{x}_{j,0} - \mathbf{x}_{j,1} \rVert^{2}_{2} - \lVert \mathbf{x}_{j,0} - \mathbf{x}_{k\neq j,1} \rVert^{2}_{2} + m) 
\label{eqn:triplet}
\end{equation}

For our implementation, the negative utterances are sampled from different speakers within the mini-batch and the sample $\mathbf{x}_k$ is selected by the hard negative mining function. This requires $M=2$ utterances from each speaker.

\newpara\noindent\textbf{Prototypical.}
Each mini-batch contains a support set $S$ and a query set $Q$. For simplicity, we will assume that the query is $M$-th utterance from every speaker. Then the prototype (or centroid) is:
\begin{equation}
\footnotesize
\mathbf{c}_{j} = \frac{1}{M-1} \sum_{m=1}^{M-1} \mathbf{x}_{j,m}
\label{eqn:proto_cent}
\end{equation}
Squared Euclidean distance is used as the distance metric as proposed by the original paper:
\begin{equation}
\footnotesize
\ms_{j,k} = \lVert \mathbf{x}_{j,M} - \mathbf{c}_k \rVert^{2}_{2}
\label{eqn:proto_dist}
\end{equation}
During training, each query example is
classified against $N$ speakers based on a softmax over distances to each speaker prototype:
\begin{equation}
\footnotesize
L_\text{P} = -\frac{1}{N} \sum_{j=1}^{N} \log
\frac{e^{\ms_{j,j}}}
{\sum_{k=1}^N e^{\ms_{j,k}}}
\label{eqn:proto_loss}
\end{equation}
Here, $\ms_{j,j}$ is the squared Euclidean distance between the query and the prototype of the same speaker from the support set. The softmax function effectively serves the purpose of hard negative mining, since the hardest negative would most affect the gradients.
The value of $M$ is typically chosen to match
the expected situation at test-time, 
{\em e.g.} $M=5+1$ for 5-shot learning, so that the prototype is composed of five different utterances.
In this way, the task in training exactly matches the task in test scenario. 

\newpara\noindent\textbf{Generalised end-to-end (GE2E).}
In GE2E training, every utterance in the batch except the query itself is used to form centroids. As a result, the centroid that is of the same class as the query is computed from one fewer utterance than centroids of other classes. They are defined as:
\begin{eqnarray}
\footnotesize
\label{eqn:ge2e_cent1}\mathbf{c}_{j} &=& \frac{1}{M} \sum_{m=1}^{M} \mathbf{x}_{j,m} \\
\label{eqn:ge2e_cent2}\mathbf{c}_{j}^{(-i)} &=& \frac{1}{M-1} \sum_{\substack{m=1\\m\neq i}}^{M} \mathbf{x}_{j,m}
\end{eqnarray}
The similarity matrix is defined as scaled cosine similarity between the embeddings and all centroids:
\begin{equation}
\footnotesize
\ms_{j,i,k} = 
\begin{cases}
w\cdot \cos(\mathbf{x}_{j,i}, \mathbf{c}_{j}^{(-i)})+b & \text{if} \quad k=j \\
w\cdot \cos(\mathbf{x}_{j,i}, \mathbf{c}_{k})+b & \text{otherwise}.
\end{cases} 
\label{eqn:ge2e_dist}
\end{equation}
where $w > 0$ and $b$ are learnable scale and bias. 
The final GE2E loss is defined as:
\begin{equation}
\footnotesize
L_\text{G} = -\frac{1}{N} \sum_{j,i} \log
\frac{e^{\ms_{j,i,j}}}
{\sum_{k=1}^N e^{\ms_{j,i,k}}}
\label{eqn:ge2e_loss}
\end{equation}

\newpara\noindent\textbf{Angular Prototypical.}
The angular prototypical loss uses the same batch formation as the original prototypical loss, reserving one utterance from every class as the query. This has advantages over GE2E-like formation since every centroid is made from the same number of utterances in the support set, therefore it is possible to exactly mimic the test scenario during training. 

We use a cosine-based similarity metric with learnable scale and bias, as in the GE2E loss.
\begin{equation}
\footnotesize
\ms_{j,k} = 
w\cdot \cos(\mathbf{x}_{j,M}, \mathbf{c}_{k})+b  
\label{eqn:angleproto_dist}
\end{equation}
Using the angular loss function introduces scale invariance, improving the robustness of objective against feature variance and demonstrating more stable convergence~\cite{wang2017deep}.

The resultant objective is the same as the original prototypical loss, Equation~\ref{eqn:proto_loss}.


\section{Experiments}
\label{sec:exp}

In this section we describe the experimental setup, which is identical across all objectives described in Section~\ref{sec:loss}.

\subsection{Input representations}
\label{subsec:input}

During training, we use a fixed length 2-second temporal segment, extracted randomly from each utterance.
Spectrograms are extracted with a hamming window of width 25ms and step 10ms. For the Thin ResNet model, the 257-dimensional raw spectrograms are used as the input to the network. For the VGG-M-40 and the Fast ResNet, 40-dimensional Mel filterbanks are used as the input.
Mean and variance normalisation (MVN) is performed by applying instance normalisation~\cite{ulyanov2016instance} to the network input.
Since the VoxCeleb dataset consists mostly of continuous speech, voice activity detection (VAD) is not used in training and testing.

\subsection{Trunk architecture}
\label{subsec:trunk}

Experiments are performed on the trunk architectures described below. The first two are identical to the models used and described in~\cite{chung2019delving}, while the last is a variation of the ResNet model to reduce computation requirement. The architectures are compared in Table~\ref{table:networks}.

\newpara\noindent\textbf{VGG-M-40.}
The VGG-M model has been proposed for image classification~\cite{Chatfield14} and adapted for speaker recognition by~\cite{Nagrani17}. The network is known for high efficiency and good classification performance. VGG-M-40 is a modification of the network proposed by~\cite{Nagrani17} to take 40-dimensional filterbanks as inputs instead of the 513-dimensional spectrogram. The temporal average pooling (TAP) layer takes the mean of the features along the time domain in order to produce utterance-level representation.

\newpara\noindent\textbf{Thin ResNet-34.}
Residual networks~\cite{He16} are widely used in image recognition and have recently been applied to speaker recognition~\cite{Chung18a,cai2018exploring,Xie19a,chung2019delving}. Thin ResNet-34 is the same as the original ResNet with 34 layers, except using only one-quarter of the channels in each residual block in order to reduce computational cost. The model only has 1.4 million parameters compared to 22 million of the standard ResNet-34. Self-attentive pooling (SAP)~\cite{cai2018exploring} is used to aggregate frame-level features into utterance-level representation while paying attention to the frames that are more informative for utterance-level speaker recognition. Thin ResNets of~\cite{cai2018exploring} and \cite{chung2019delving} differ slightly in their implementation details, but in our experiments we use that of~\cite{chung2019delving}.

\newpara\noindent\textbf{Fast ResNet-34.}
The number and size of filters are identical to the Thin ResNets of~\cite{cai2018exploring,chung2019delving}, but the input dimensions are smaller than~\cite{chung2019delving} and the strides are earlier than~\cite{cai2018exploring} in order to reduce computational requirements. Due to space constraints, the exact specification can be found in the accompanying code. The performance is on par with both Thin ResNet models, while the computation cost is less than half of those models.

\begin{table}[ht]
\centering
\begin{tabular}{ |l|r|r|r| }
\hline
\textbf{Network} & \textbf{Params}  & \textbf{MACs}   \\ \hline\hline
{\bf VGG-M-40}~\cite{chung2019delving}   & 4.0M & 0.53G \\ \hline
{\bf Thin ResNet-34}~\cite{chung2019delving}   & 1.4M & 0.99G  \\ \hline
{\bf Thin ResNet-34}~\cite{cai2018exploring}   & 1.4M & 0.93G  \\ \hline
{\bf Fast ResNet-34}                    & 1.4M & 0.45G \\ \hline
\end{tabular}
\vspace{5pt}
\caption{Network statistics. Multiply–accumulate operations (MACs) are measured for a 2-second input.}
\label{table:networks}
\vspace{-20pt}
\end{table}

\begin{table*}[h!] 
\begin{center}
\footnotesize
\begin{tabular}{ |l  | l |r|r|r|} 
 \hline
 \textbf{~Objective} &
 \textbf{Hyperparameters} &
 \textbf{~~~~~~VGG-M-40} &
 \textbf{Thin ResNet-34} &
 \textbf{Fast ResNet-34} 
 \\ \hline\hline
 
 ~Softmax & - & $10.14\pm0.20$ & $5.82\pm0.47$  & $6.46\pm0.06$  \\ \hline
   \multirow{12}{90pt}{~AM-Softmax~\cite{wang2018additive}} & 
           $m=0.1$, $s=15$  & $4.86\pm0.14$ &   $2.81\pm0.08$   & $2.77\pm0.03$  \\
           & $m=0.2$, $s=15$ & $5.14\pm0.13$ &   $2.85\pm0.07$  & $3.05\pm0.03$  \\
           & $m=0.3$, $s=15$ & $5.24\pm0.08$ &   $3.08\pm0.05$  & $3.08\pm0.08$  \\
          & $m=0.4$, $s=15$ & $5.22\pm0.15$ &   $3.09\pm0.06$   & $3.25\pm0.09$  \\
           & $m=0.1$, $s=30$ & $\mathbf{4.76\pm0.10}$ &  $2.59\pm0.09$ & $\mathbf{2.41\pm0.01}$ \\
           & $m=0.2$, $s=30$ & $4.88\pm0.03$ &   $\mathbf{2.40\pm0.07}$ & $2.43\pm0.05$   \\
           & $m=0.3$, $s=30$ & $5.19\pm0.08$ &   $2.71\pm0.10$  & $2.52\pm0.04$  \\
          & $m=0.4$, $s=30$ & $5.35\pm0.06$ &   $2.81\pm0.10$   & $2.67\pm0.05$  \\
           & $m=0.1$, $s=50$ & $5.45\pm0.06$ &  $2.99\pm0.04$   & $2.73\pm0.07$   \\
           & $m=0.2$, $s=50$ & $5.28\pm0.07$ &   $2.60\pm0.10$  & $2.51\pm0.01$   \\
           & $m=0.3$, $s=50$ & $5.62\pm0.09$ &   $2.80\pm0.09$  & $2.53\pm0.06$  \\
          & $m=0.4$, $s=50$ & $5.91\pm0.12$ &   $2.96\pm0.08$   & $2.69\pm0.07$  \\\hline
   \multirow{12}{90pt}{~AAM-Softmax~\cite{deng2019arcface}} & 
             $m=0.1$, $s=15$                    & $4.81\pm0.03$ &       $2.78\pm0.04$  & $2.80\pm0.11$   \\
             & $m=0.2$, $s=15$                    & $4.88\pm0.08$ &     $2.88\pm0.09$  & $2.98\pm0.05$  \\
             & $m=0.3$, $s=15$                    & $14.90\pm0.16$   &  $3.16\pm0.05$  & $14.98\pm0.20$   \\
           & ~~~~~~~~~~$\rightarrow$ Curriculum    & $5.00\pm0.05$ &    $2.91\pm0.08$  & $3.04\pm0.06$  \\
           
           & $m=0.1$, $s=30$                    & $4.67\pm0.06$ &       $2.60\pm0.07$  & $2.48\pm0.02$  \\
           & $m=0.2$, $s=30$            &  $\mathbf{4.64\pm0.04}$ &     $\mathbf{2.36\pm0.04}$ &   $2.38\pm0.01$   \\
           & $m=0.3$, $s=30$                    & $13.25\pm0.07$   &    $10.55\pm0.33$ & $11.35\pm0.18$     \\
           & ~~~~~~~~~~$\rightarrow$ Curriculum    & $4.69\pm0.02$ &    $2.39\pm0.05$  & $\mathbf{2.37\pm0.02}$  \\
           
           & $m=0.1$, $s=50$                    & $5.27\pm0.03$ &       $2.88\pm0.05$  & $2.71\pm0.07$    \\
           & $m=0.2$, $s=50$                    & $4.96\pm0.03$ &       $2.50\pm0.05$  & $2.49\pm0.04$   \\
           & $m=0.3$, $s=50$                    & $10.42\pm0.12$ &      $8.79\pm0.21$  & $9.49\pm0.25$  \\
           & ~~~~~~~~~~$\rightarrow$ Curriculum    & $4.86\pm0.11$ &    $2.41\pm0.08$  & $2.42\pm0.06$ \\\hline\hline
           
   \multirow{4}{90pt}{~Triplet~\cite{Schroff15}}  &
             $m=0.1$,~~CHNM   & $4.86\pm0.15$ &             $\mathbf{2.53\pm0.10}$   & $2.73\pm0.03$ \\
           & $m=0.2$,~~CHNM   & $\mathbf{4.67\pm0.06}$ &    $2.60\pm0.02$           &  $\mathbf{2.71\pm0.06}$  \\
           & $m=0.3$,~~CHNM   & $4.84\pm0.13$ &             $2.66\pm0.03$           & $2.85\pm0.04$ \\
           & $m=0.4$,~~CHNM   & $4.84\pm0.08$&              $2.76\pm0.10$           & $2.96\pm0.07$ \\\hline
    \multirow{5}{90pt}{~GE2E~\cite{wan2018generalized}}  &
             $M=2$  & $4.60\pm0.04$ &               $2.56\pm0.08$           & $2.51\pm0.07$ \\
           & $M=3$  & $\mathbf{4.40\pm0.08}$ &      $\mathbf{2.52\pm0.07}$  & $\mathbf{2.37\pm0.10}$  \\
           & $M=4$  & $4.49\pm0.05$ &               $2.59\pm0.12$           & $2.59\pm0.08$ \\
           & $M=5$  & $4.69\pm0.09$&                $2.78\pm0.09$           & $2.66\pm0.02$ \\
           & $M=10$  & $5.53\pm0.04$ &              $3.68\pm0.08$           & $3.55\pm0.05$    \\\hline
   \multirow{4}{90pt}{~Prototypical~\cite{snell2017prototypical}}  &
               $M=2$  & $\mathbf{4.59\pm0.02}$ &   $\mathbf{2.34\pm0.08}$  &   $\mathbf{2.32\pm0.02}$  \\
           & $M=3$  & $4.73\pm0.11$ &               $2.54\pm0.07$   & $2.39\pm0.05$ \\
           & $M=4$  & $4.99\pm0.19$ &               $2.83\pm0.04$   & $2.89\pm0.04$ \\
           & $M=5$  & $5.34\pm0.03$ &               $3.33\pm0.11$   & $3.21\pm0.01$ \\\hline

    \multirow{4}{90pt}{\bf Angular Prototypical}  &
      $M=2$ & $\mathbf{4.29\pm0.07}$ & $\mathbf{2.21\pm0.03}$ & $\mathbf{2.22\pm0.05}$ \\ 
    & $M=3$ & $4.30\pm0.05$ & $2.45\pm0.07$ & $2.40\pm0.04$ \\ 
    & $M=4$ & $4.53\pm0.03$ & $2.75\pm0.06$ & $2.60\pm0.02$ \\ 
    & $M=5$ & $4.73\pm0.01$ & $3.00\pm0.11$ & $2.90\pm0.11$ \\ \hline
\end{tabular}
\end{center}
\caption{
Equal Error Rates (EER, \%) on the VoxCeleb1 test set. We report the mean and standard deviation of the repeated experiments. {\bf CHNM}: Curriculum Hard Negative Mining.
}
\label{tab:results}
\end{table*}

\begin{table*}[h!] 
\begin{center}
\footnotesize
\begin{tabular}{ |l  | l | r |r|r|r|} 
 \hline
 \textbf{Objective} & \textbf{Hyperparameters} &
 \textbf{200} &
 \textbf{400} &
 \textbf{600} &
 \textbf{800}
 \\ \hline\hline
 
 AM-Softmax &$m=0.2$, $s=30$                & $2.40\pm0.07$ & $2.53\pm0.08$ & $2.49\pm0.11$ & $2.57\pm0.07$ \\ \hline
 Prototypical & $M=2$                       & $2.42\pm0.04$ & $2.40\pm0.07$ & $2.34\pm0.05$ & $2.34\pm0.08$ \\ \hline
 {\bf Angular Prototypical}~~~ & $M=2$         & $2.37\pm0.07$ & $2.31\pm0.05$ & $2.32\pm0.09$ & $2.21\pm0.03$ \\ \hline
\end{tabular}
\end{center}
\caption{
Effect of training batch size on test performance. Equal Error Rates (EER, \%) using the Thin ResNet-34 architecture on the VoxCeleb1 test set. We report the mean and standard deviation of the repeated experiments.
}
\label{tab:batchsize}
\end{table*}

\subsection{Implementation details}
\label{subsec:impl}

\newpara\noindent\textbf{Datasets.}
The network is trained on the development set of VoxCeleb2~\cite{Chung18a} and evaluated on test set of VoxCeleb1~\cite{Nagrani17}.
Note that the development set of
VoxCeleb2 is completely disjoint from the VoxCeleb1
dataset (\textit{i.e.} no speakers in common).

\newpara\noindent\textbf{Training.}
Our implementation is based on the PyTorch framework~\cite{paszke2019pytorch} and trained on the NAVER Smart Machine Learning (NSML) platform~\cite{sung2017nsml}.
The models are trained using a NVIDIA V100 GPU with 32GB memory for $500$ epochs. 
For each epoch, we randomly sample a maximum of 100 utterances from each of the 5,994 identities to reduce class imbalance.
We use the Adam optimizer with an initial learning rate of $0.001$ decreasing by $5\%$ every 10 epochs. 
For metric learning objectives, we use the largest batch size that fits on a GPU. For classification objectives, we use a fixed batch size of 200.
The training takes approximately one day for the VGG-M-40 model, two days for the Fast ResNet model and five days for the Thin ResNet model.

All experiments were repeated independently three times in order to minimise the effect of random initialisation, and we report mean and standard deviation of the experiments. 

\newpara\noindent\textbf{Data augmentation.}
No data augmentation is performed during training, apart from the random sampling.

\newpara\noindent\textbf{Curriculum learning.}
The AAM-Softmax loss function demonstrates unstable convergence from random initialisation with larger values of $m$ such as $0.3$. Therefore, we start training the model with $m=0.1$ and increase it to $m=0.3$ after 100 epochs. This strategy is labelled {\em Curriculum} in Table~\ref{tab:results}. 

Similarly, the triplet loss can cause models to diverge if the triplets are too difficult early in the training. We only enable hard negative mining after 100 epochs, at which point the network only sees the most difficult 1\% of the negatives.


\subsection{Evaluation}
\label{subsec:eval}

\newpara\noindent\textbf{Evaluation protocol.}
The trained networks are evaluated on the VoxCeleb1 test set. We sample ten 4-second temporal crops at regular intervals from each test segment, and compute the similarities between all possible combinations
($10 \times 10 = 100$) from every pair of segments. The mean of the 100 similarities is used as the score. This protocol is in line with that used by~\cite{Chung18a,chung2019delving}.

\newpara\noindent\textbf{Results.}
The results are given in Table~\ref{tab:results}. It can be seen that the performance of networks trained with AM-Softmax and AAM-Softmax loss functions can be very sensitive to the value of margin and scale set during training. We iterate over many combinations of $m$ and $s$ to find the optimal value. The model trained with the most common setting (AM-Softmax with $m=0.3$ and $s=30$) is outperformed by the vanilla triplet loss.

Generalised end-to-end and prototypical losses show improvements over the triplet loss by using multiple negatives in training. The prototypical networks perform best when the value of $M$ matches the test scenario, removing the necessity for hyperparameter optimisation. The performance of the model trained with the proposed angular objective exceeds that of all classification-based and metric learning methods.

There are a substantial number of recent works on the VoxCeleb2 dataset, but we do not compare to these in the table, since the goal of this work is to compare the performance of different loss functions under identical conditions. However, we are unaware of any work that outperforms our method with a similar number of network parameters.

\newpara\noindent\textbf{Batch size.}
The effect of batch size on various loss functions is shown in Table~\ref{tab:batchsize}. We observe that a bigger batch size has a positive effect on performance for metric learning methods, which can be explained by the ability to sample harder negatives within the batch. We make no such observation for the network trained with classification loss.


\section{Conclusions}

In this paper, we have presented a case for metric learning in speaker recognition. Our extensive experiments indicate that the GE2E and prototypical networks show superior performance to the popular classification-based methods. We also propose an angular variant of the prototypical networks that outperforms all existing training functions. Finally, we release a flexible PyTorch trainer for large-scale speaker recognition that can be used to facilitate further research in the field.

\clearpage
\raggedbottom
\bibliographystyle{IEEEtran}
\bibliography{shortstrings,mybib,vgg_local,vgg_other}
\end{document}